\documentclass[12pt]{article}
\usepackage{latexsym}
\newcommand{\beq}{\begin{eqnarray}}
\newcommand{\eeq}{\end{eqnarray}}
\begin{document}
\title{Density of states for Bose-Einstein condensation in harmonic  
oscillator potentials}
\author{Klaus Kirsten    \cite{kk}\\University of Leipzig\\
Institute of Theoretical Physics\\
Augustusplatz 10, 
04109 Leipzig\\
\\and\\
\\
David J. Toms   \cite{djt}          \\
Department of Physics\\ University of Newcastle Upon Tyne,\\
Newcastle Upon Tyne, U. K. NE1 7RU}

\date{\today}
\maketitle

\begin{abstract}
We discuss how it is possible to obtain a reliable approximation for the  
density of states for a system of particles in an anisotropic harmonic  
oscillator potential. A direct application of the result to study  
Bose-Einstein condensation of atomic gases in a potential trap can be  
given. In contrast to a previous study, our method involves only  
analytic calculations.
\end{abstract}
\eject

Recent experiments on cold alkali gases \cite{rub,lith,sod} have given  
the best evidence so far for Bose-Einstein condensation (BEC). The  
experiments demonstrate that below a characteristic temperature on the  
order of a few degrees microKelvin several thousand atoms are found in  
the ground state. Because the gases are dilute, as a first approximation  
interatomic interactions should be negligible. However the free boson  
gas model as described in the standard textbooks \cite{texts} is not a  
good theoretical model for these atomic systems due to the effects of  
the confining potential traps used in the experiments. To a good  
approximation the magnetic traps can be modelled by harmonic oscillator  
potentials. There have been several studies of BEC for a system of  
particles in a harmonic oscillator potential  
\cite{oldone,MITcrowd,GHPLA,GHNat,HaugRav,KKDJT1,KKDJT2}.

If we consider a system of particles in an external anisotropic harmonic  
oscillator potential characterized by angular frequencies  
$\omega_1,\omega_2$ and $\omega_3$, then the energy levels are (with  
$\hbar=1$)
\beq
E_{n_1n_2n_3}=(n_1+\frac{1}{2})\omega_1+(n_1+\frac{1}{2})\omega_2
+(n_1+\frac{1}{2})\omega_3\;,\label{eq1}
\eeq
where $n_{1,2,3}=0,1,2,\ldots$. If we describe the system using the  
grand canonical ensemble (see Ref.~\cite{GHprep} for a treatment using  
the microcanonical ensemble), then the thermodynamics may be studied  
from the grand potential (see Pathria in Ref.~\cite{texts})
\beq
q=-\sum_{n_1,n_2,n_3}\ln\Big\lbrack1-e^{-\beta(E_{n_1n_2n_3}-\mu)}
\Big\rbrack\;.\label{eq2}
\eeq
The number of particles is
\beq
N=\sum_{n_1,n_2,n_3}\Big\lbrack e^{\beta(E_{n_1n_2n_3}-\mu)}
-1\Big\rbrack^{-1}\;.\label{eq3}
\eeq
It is not possible to evaluate the sums in (\ref{eq2}) or (\ref{eq3})  
directly in closed form. There have been two main approaches to try and  
obtain analytic approximations for such sums.

One method is to argue that for situations of experimental interest  
since the temperature $T$ is of the order a few microkelvin and  
$\omega_i$ is of the order a few hundred Hz, we have $\beta\omega_i\ll  
1$. It is therefore possible to approximate the sums over discrete  
energy levels in expressions such as (\ref{eq2}) and (\ref{eq3}) by  
integrals. A crucial feature in obtaining a reliable approximation is to  
use an appropriate density of states. A beautiful and careful account of  
the correct way to analyze this problem was given by Grossmann and  
Holthaus \cite{GHNat}, and used in Ref.~\cite{GHPLA}. These authors  
noted that the density of states included an additional term beyond that  
found in the unconfined Bose gas. They postulated that the density of  
states was given approximately by
\beq
\rho(E)=\frac{1}{2}\frac{E^2}{\Omega^3}+\gamma\frac{E}{\Omega^2}
\;,\label{eq4}
\eeq
where $\Omega=(\omega_1\omega_2\omega_3)^{1/3}$. If only the first term  
on the right hand side is included as in Refs.~\cite{MITcrowd,HaugRav}  
it leads to an unphysical discontinuous behaviour in the specific heat.  
The second term on the right hand side of (\ref{eq4}) involves a  
constant $\gamma$ which for the isotropic harmonic oscillator is easily  
seen to be 3/2, but which Grossmann and Holthaus had to evaluate  
numerically in the anisotropic case.

A different method for studying (\ref{eq2}) and (\ref{eq3}) was used by  
the present authors \cite{KKDJT1,KKDJT2}. Instead of converting the sums  
directly into integrals using a density of states factor, a contour  
integral representation was found for (\ref{eq2}) and (\ref{eq3}) with  
the sums being performed exactly. By deforming the integration contour  
in an appropriate way asymptotic expansions were found for $q,N$ and  
other thermodynamic quantities. These calculations, in contrast with  
those of Refs.~\cite{GHPLA,GHNat}, were entirely analytic. Given that  
the approaches of Refs.~\cite{GHPLA,GHNat} and \cite{KKDJT1,KKDJT2} both  
give good agreement with a direct numerical evaluation of the sums  
(\ref{eq2}) and (\ref{eq3}), this indicates that by comparing the two  
sets of results for say the particle number, it is possible to extract  
an analytic result for $\gamma$ in the anisotropic case. As this is  
rather a roundabout way of proceeding, in the present paper we wish to  
describe a more direct way of evaluating the density of states, the  
result of which leads to an exact equality between the approach of  
Grossmann and Holthaus \cite{GHPLA,GHNat} and that of our earlier work  
\cite{KKDJT1,KKDJT2}.

The basic idea involves the use of the asymptotic high-temperature
expansion of the partition function ${\cal Z} (\beta) $
for the simple harmonic oscillator. Generally, if we define  
$E_N$ to be the eigenvalues, labelled by some index $N$
($E_N\leq E_{N+1}$), of  
some self-adjoint differential operator, we may form the sum
\beq
{\cal Z} (\beta )=\sum_Ne^{-\beta E      _N}
 =\int_b^{\infty}e^{-\beta E} 
\rho_d (E) dE               \;    , \label{eq5}
\eeq
with $0<b<E_0$ and $\rho_d (E) =\sum_N \delta (E-E_N)$ is the density of
states. Approximating $\rho_d (E)$ by a smooth function $\rho_s (E)$, 
a refinement of Karamatas theorem states 
\cite{Brownell,BH}, that if 
\beq
{\cal Z} (\beta )=\int_b^{\infty}e^{-\beta E} 
\rho_s (E) dE \simeq  
\sum_{i=1}^{k}c_i   \beta^{-r_i}+{\cal O}\Big(\beta^{-r_{k+1}}\Big)
\;,\label{eq6}
\eeq
for $0<\beta \leq C$ for some real $C>0$ and $r_{k+1} < r_k <
\cdots <r_1$, then
for $E>b$ 
\beq
\rho_s (E) = \sum_{i=1}^k 
\frac{c_i}{\Gamma (r_i)} E^{r_i -1} 
+{\cal O} (E^{r_{k+1}-1} \ln E) , 
\label{eq7}
\eeq
where $1/\Gamma (r_i )$ is defined to be zero if $r_i $ is a non-positive
integer.
By virtue of this theorem, an asymptotic expansion for the Laplace
transform of the density of states is carried over into an 
expansion for the density of states itself. 
Whereas the Laplace transform method can only be used if $r_i >0$, the above
theorem extends to all values of $r_i$.
An excellent and accessible discussion of this theorem with
extensive references and applications is the book by Baltes and Hilf
\cite{BH}.

The coefficients $c_i$ in (\ref{eq6}) are known for a wide class of  
operators on general Riemannian manifolds with a variety of boundary  
conditions
\cite{Minaketal}.
However in the case of the anisotropic oscillator a direct  
evaluation of the $c_i$ is very simple. For the eigenvalues $E      _N$  
in (\ref{eq5}) we take the energy levels given in (\ref{eq1}). However,  
because we are interested in BEC where the ground state plays a key  
role, we single out the ground state for special treatment. If the  
ground state is separated off from the sums in (\ref{eq2}) and  
(\ref{eq3}), then the partition sum for determining the density of  
states is 
\beq
{\cal Z} (\beta) =  
\sum_{n_1,n_2,n_3=0}^{\infty}e^{-\beta (n_1\omega_1+n_2\omega_2+ 
n_3\omega_3)}
=\prod_{i=1}^{3}\Big(1-e^{-\beta \omega_i}\Big)^{-1}\;.\label{eq9}
\eeq
The case of the harmonic oscillator potential is particularly simple  
since the sum in (\ref{eq5}) involves only geometric series. It is now  
straightforward to expand (\ref{eq9}) in powers of $t$ using
\beq
\Big(1-e^{-x}\Big)^{-1}=\frac{1}{x}+\frac{1}{2}+\frac{1}{12}x
+{\cal O}(x^3)\;.\label{eq10}
\eeq
(The coefficients appearing in the expansion are just the Bernoulli  
numbers.) It is easily found that
\beq
{\cal Z} (\beta ) \simeq c_1\beta 
^{-3}+c_2\beta
^{-2}+c_3\beta
^{-1}+\cdots\;,\label{eq11}
\eeq
where
\beq
c_1&=&(\omega_1\omega_2\omega_3)^{-1}\label{eq12}\\
c_2&=&\frac{1}{2}\left(\frac{1}{\omega_1\omega_2}+
\frac{1}{\omega_2\omega_3}+\frac{1}{\omega_1\omega_3}\right)\label{eq13} 
\\
c_3&=&\frac{1}{12}\left(\frac{\omega_3}{\omega_1\omega_2}+
\frac{\omega_2}{\omega_1\omega_3}+\frac{\omega_1}{\omega_2\omega_3}+
\frac{3}{\omega_1}+\frac{3}{\omega_2}+\frac{3}{\omega_3}\right)
\;.\label{eq14}
\eeq
Using (\ref{eq7}) the density of states is found to be
\beq
\rho(E)\simeq\frac{1}{2}c_1E^2+c_2E+c_3\;.\label{eq15}
\eeq

Comparison of our result with the ansatz of Grossmann and Holthaus in  
(\ref{eq4}) shows that the coefficient of $E^2$ is correct, and in  
addition fixes their constant $\gamma$ in terms of the oscillator  
frequencies. There is an additional constant term in the density of  
states which will make a small correction to the results of  
Refs.~\cite{GHPLA,GHNat}. 
However, slightly below the critical temperature $T_c$ this conversion
error is expected to be of the order of or less than the 
discretization error and it may be consistently neglected excluding
a temperature interval around $T_c$ of the order $N^{-1/3}$ from the 
analysis \cite{GHNat}.
In the special case of the isotropic harmonic  
oscillator with $\omega_1=\omega_2=\omega_3=\Omega$ it can be seen that  
our result reduces to $c_1=\Omega^{-3}, c_2=\frac{3}{2}\Omega^{-2}$ and  
$c_3=\Omega^{-1}$. This is also in agreement with the results of  
Ref.~\cite{GHNat}.

Use of the modified form of the density of states (\ref{eq15}) to  
discuss BEC is basically identical to that found in  
Refs.~\cite{GHPLA,GHNat}, and there is no need to repeat the analysis  
here. The coefficients (\ref{eq12}--\ref{eq14}) also enter the  
asymptotic expansions of Refs.~\cite{KKDJT1,KKDJT2}, and it is now  
straightforward to establish a direct connection between the results of  
Refs.~\cite{GHPLA,GHNat} and those of Refs.~\cite{KKDJT1,KKDJT2}.

After this work was completed we became aware of another systematic  
treatment of Bose-Einstein condensation in harmonic oscillator traps  
\cite{Rav}. This reference uses the Euler-Maclaurin summation technique  
to analyze the harmonic oscillator sums. Detailed analysis is given for  
the isotropic harmonic oscillator which appears to agree with our  
earlier work \cite{KKDJT1,KKDJT2}. A result for the particle number is  
given for the case of the anisotropic oscillator which is also in  
agreement with the results of our evaluation \cite{KKDJT2}, as well as  
the result found by using the first two terms in the density of states  
approach given in the present paper.

In conclusion, we have shown how to calculate the density of states  
factor for a system of particles in an anisotropic 
harmonic oscillator potential.  
This means that an analysis similar to that of Grossmann and Holthaus  
\cite{GHPLA,GHNat} can be performed in an entirely analytic manner  
without any resort to numerical computation. The method we have outlined  
can be extended to many other situations of interest, and we will  
present a more complete discussion elsewhere.

\vspace{1cm}
\noindent{\bf Acknowledgements}

D.~J.~T. would like to thank J.~D.~Smith for helpful discussions.


\begin{thebibliography}{99}
\bibitem[\dag]{kk}
{\tt kirsten@tph100.physik.uni-leipzig.de}.
\bibitem[\ddag]{djt}
{\tt d.j.toms@newcastle.ac.uk}.
\bibitem{rub}
M.~H.~Anderson, J.~R.~Ensher, M.~R.~Matthews, C.~E.~Wieman, and  
E.~A.~Cornell, Science {\bf 269}, 198 (1995).
\bibitem{lith}
C.~C.~Bradley, C.~A.~Sackett, J.~J.~Tollett, and R.~G.~Hulet, Phys. Rev.  
Lett. {\bf 75}, 1687 (1995).
\bibitem{sod}
K.~B.~Davis, M.-O.~Mewes, M.~R.~Andrews, N.~J.~van Druten, D.~S.~Durfee,  
D.~M.~Kurn, and W.~Ketterle, Phys. Rev. Lett. {\bf 75}, 3969 (1995).
\bibitem{texts}
See for example F. London, {\em Superfluids II\/}, (John Wiley, New  
York, 1954), K. Huang, {\em Statistical Mechanics\/} (John Wiley, New  
York, 1987) or R.~K.~Pathria, {\em Statistical Mechanics\/} (Pergammon,  
Oxford, 1972).
\bibitem{oldone}
S.~R.~de Groot, G.~J.~Hooyman, and C.~A.~ten Seldam, Proc. Roy. Soc.  
(London) {\bf A 203}, 266 (1950).
\bibitem{MITcrowd}
V.~Bagnato, D.~E.~Pritchard, and D.~Kleppner, Phys. Rev. A {\bf 35},  
4354 (1987).
\bibitem{GHPLA}
S.~Grossmann and M.~Holthaus, Phys. Lett. A {\bf 208}, 188 (1995).
\bibitem{GHNat}
S.~Grossmann and M.~Holthaus, Zeit. f\"{u}r Naturforschung {\bf 50 a},  
921 (1995).
\bibitem{HaugRav}
H.~Haugerud and F.~Ravndal (unpublished), cond-mat/9509041.
\bibitem{KKDJT1}
K.~Kirsten and D.~J.~Toms, Bose-Einstein condensation of atomic gases in  
a harmonic oscillator confining potential trap, preprint, Feb. 1996,  
cond-mat/9604031.
\bibitem{KKDJT2}
K.~Kirsten and D.~J.~Toms, Bose-Einstein condensation of atomic gases in  
general harmonic oscillator confining potential traps, preprint, March  
1996, to appear in Phys.~Rev.~A.
\bibitem{GHprep}
S.~Grossmann and M.~Holthaus, Microcanonical fluctuations of a Bose  
system's ground state occupation number, preprint, Feb. 1996.
\bibitem{Brownell}
F.~H.~Brownell, Pacific J. Math. {\bf 5}, 483 (1955); J. Math. Mech.  
{\bf 6}, 119 (1957).
\bibitem{BH}
H.~P.~Baltes and E.~R.~Hilf, {\em Spectra of Finite Systems\/}  
(Bibliographisches Institut Mannheim, Zurich, 1976).
\bibitem{Minaketal}
S.~Minakshisundaram, Can. J. Math. {\bf 1}, 320 (1949);  
S.~Minakshisundaram and \AA.~Pleijel, Can. J. Math. {\bf 1}, 242 (1949);  
B.~S.~DeWitt, {\em Dynamical Theory of Groups and Fields\/} (Gordon and  
Breach, New York, 1965); 
P.~B.~Gilkey, {\em Invariance theory, the heat equation and the 
Atiyah-Singer index theorem, 2nd.~Edn.} (CTC Press, Boca Raton, 1995). 
\bibitem{Rav}
H.~Haugerud, T.~Haugset, and F.~Ravndal, A more accurate analysis of  
Bose-Einstein condensation in harmonic traps, preprint, May 1996,  
cond-mat/9605100.

\end{thebibliography}
\end{document}